\documentclass[twocolumn,preprintnumbers,amsmath,amssymb,prl,superscriptaddress]{revtex4}

\usepackage[dvipsnames]{xcolor}
\usepackage{graphicx}
\usepackage{bbm}

\newcommand{\SFA}{SrFe$_2$As$_2$}
\newcommand{\BFA}{BaFe$_2$As$_2$}

\newcommand{\be}{\begin{eqnarray}}
\newcommand{\ee}{\end{eqnarray}}

\begin{document}

\title{Impact of the first order antiferromagnetic phase transition on the paramagnetic spin excitations and nematic phase of SrFe$_2$As$_2$}

\author{David W. Tam}
\thanks{These authors made equal contributions to this work.}
\affiliation{Department of Physics and Astronomy, Rice University, Houston, Texas 77005, USA}

\author{Weiyi Wang}
\thanks{These authors made equal contributions to this work.}
\affiliation{Department of Physics and Astronomy, Rice University, Houston, Texas 77005, USA}

\author{Li Zhang}
\email{lzhang@cjlu.edu.cn}
\affiliation{Department of Physics, China Jiliang University, Hangzhou 310018, China,}
\affiliation{Department of Physics and Astronomy, Rice University, Houston, Texas 77005, USA}

\author{Yu Song}
\author{Rui Zhang}
\author{Scott V. Carr}
\affiliation{Department of Physics and Astronomy, Rice University, Houston, Texas 77005, USA}

\author{H. C. Walker}
\author{Toby G. Perring}
\affiliation{ISIS Facility, Rutherford Appleton Laboratory, Chilton, Didcot, Oxfordshire OX11 0QX, UK}

\author{D. T. Adroja}
\affiliation{ISIS Facility, Rutherford Appleton Laboratory, Chilton, Didcot, Oxfordshire OX11 0QX, UK}
\affiliation{Highly Correlated Matter Research Group, Physics Department, University of Johannesburg, 
P.O. Box 524, Auckland Park 2006, South Africa} 

\author{Pengcheng Dai}
\email{pdai@rice.edu}
\affiliation{Department of Physics and Astronomy, Rice University, Houston, Texas 77005, USA}

\date{\today}

\begin{abstract}
Understanding the nature of the electronic nematic phase in iron pnictide superconductors is important for elucidating its impact on high-temperature superconductivity.  
Here we use transport and inelastic neutron scattering to study spin excitations and in-plane resistivity anisotropy in uniaxial pressure detwinned BaFe$_2$As$_2$ and SrFe$_2$As$_2$, the parent compounds of iron pnictide superconductors.
While BaFe$_2$As$_2$ exhibits weakly first order tetragonal-to-orthorhombic structural and antiferromagnetic (AF) phase transitions below $T_s > T_N\approx 138$ K, SrFe$_2$As$_2$ has strongly coupled first order structural and AF transitions below $T_s= T_N\approx 210$ K. 
We find that the direct signatures of the nematic phase
 persist to lower temperatures above the phase transition in the case of 
SrFe$_2$As$_2$ compared to BaFe$_2$As$_2$.
Our findings support the conclusion that the strongly first-order nature of the magnetic transition
in SrFe$_2$As$_2$ weakens the nematic phase and resistivity anisotropy in the system. 
\end{abstract}

\maketitle

\section{Introduction}

\begin{figure}
\includegraphics[scale=.45]{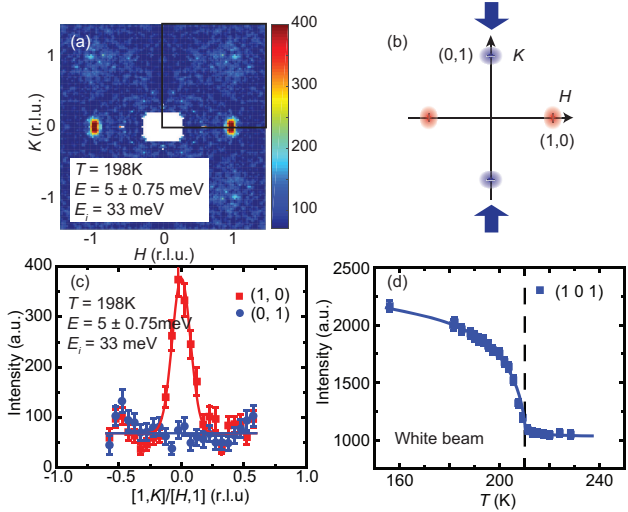}
\caption{Summary of the inelastic neutron scattering results on uniaxial pressure detwinned SrFe$_2$As$_2$.
(a) Spin waves at an energy transfer $E=5\pm 0.75$ meV from a $\sim$100\% detwinned SrFe$_2$As$_2$ below $T_N$, where magnetic intensity are at $\mathbf{Q}_{\rm AF} = (\pm1, 0)$ and absent at $(0, \pm1)$.
(b) Reciprocal space of SrFe$_2$As$_2$ with twin domains. The blue and red dots mark the magnetic Bragg peak positions for the two twin domains. When uniaxial pressure is applied along the $b$-axis direction, only Bragg peaks and spin waves from the red domain are present. 
(c) Cuts of $E=5\pm 0.75$ meV spin waves along the red and blue positions in reciprocal space at $T=198$ K ($<T_{\rm N}$). The absence of magnetic scattering at $(0, \pm1)$ indicates that the sample is essentially 100\% detwinned.
(d) Temperature dependence of the magnetic Bragg peak's intensity at $\mathbf{Q}_{\rm AF} = (\pm1, 0, 1)$. Note that $T_N$ is increased under uniaxial pressure.
}
\label{fig1n}
\end{figure}

The parent compounds of iron-based superconductors such as BaFe$_2$As$_2$ and SrFe$_2$As$_2$ exhibit antiferromagnetic (AF) order below the phase transition temperature $T_{\rm N}$ \cite{scalapinormp,stewart,dairmp}. 
At temperatures at or slightly above $T_{\rm N}$, these materials also exhibit a tetragonal-to-orthorhombic structural transition at $T_{\rm s}$, where the underlying lattice changes from having four-fold ($C_4$) above $T_{\rm s}$ to two-fold ($C_2$) rotational symmetry below $T_{\rm s}$ \cite{stewart,dairmp}.
In the temperature regime below $T_{\rm s}$ and above $T_{\rm N}$, an electronic nematic phase, which breaks the orientational but not the translational symmetry of the underlying lattice \cite{Fradkin}, has been predicted \cite{fang08,CXu}.
As the nematic phase and associated fluctuations can act to enhance electron Cooper pairing for superconductivity \cite{Tmaier,Metlitski,SLederer,SLederer17} and is expected to play an important role in iron pinctides \cite{RMFernandes2014}, it is important to elucidate its microscopic origin.
However, in the unstrained state, BaFe$_2$As$_2$ and SrFe$_2$As$_2$ form twinned domains below $T_{\rm s}$, making it impossible for a bulk probe to determine the intrinsic electronic properties of the individual domains or the associated nematic fluctuations.
By applying uniaxial pressure along one of the orthorhombic lattice directions, one can detwin BaFe$_2$As$_2$ single crystals and therefore measure the intrinsic electronic anisotropy present in the orthorhombic phase \cite{Fisher11}.
When the material is completely detwinned, the magnetic Bragg peaks from the collinear AF order below $T_{\rm N}$ will appear at the in-plane $\mathbf{Q}_{\rm AF} = (\pm1, 0)$ wave vectors in reciprocal space, with no observable peaks at $(0, \pm1)$ from the extinguished domain [Figs. \ref{fig1n}(a) and \ref{fig1n}(b)] \cite{qhuang,mgkim,JZhao2008}.
As a result of this technique, one can then examine the material at temperatures above $T_{\rm N}$ to elucidate the microscopic nature of the nematic phase.

From the temperature dependence of the in-plane resistivity anisotropy measured on uniaxial pressure detwinned BaFe$_2$As$_2$, the electronic nematic phase has been identified to persist to a characteristic temperature $T^\ast$ higher than the expected nematic ordering temperature $T_{\rm s}$ \cite{JHChu2010,JHChu2012}.
In previous transport and inelastic neutron scattering studies of uniaxial pressure detwinned BaFe$_2$As$_2$ \cite{Dhital12,Lu14,HRMAN15,Wenliang,HRMan18}, resistivity anisotropy in the paramagnetic phase above $T_s$ is found to be associated with anisotropy in spin excitations between the AF wavevector $\mathbf{Q}_{\rm AF} = (\pm1, 0)$ and the disallowed wavevector $\mathbf{Q}=(0, \pm1)$, thus suggesting that the nematic phase is driven by magnetism \cite{fernandes11} instead of orbital ordering \cite{lee,kruger,lv}.
For BaFe$_2$As$_2$, which has separate weakly first order magnetic and second order structural phase transitions ($T_{\rm s}>T_{\rm N}$ by $\sim$0.75 K) \cite{mgkim}, one would expect that critical spin fluctuations from the AF phase transition extend to temperatures well above $T_{\rm N}$.
On the other hand, for SrFe$_2$As$_2$, which has strongly coupled first order magnetic and structural phase transitions ($T_{\rm s}=T_{\rm N}$) \cite{Krellner,Jesche}, there should not be much critical scattering above $T_{\rm N}$.
If nematic fluctuations in the paramagnetic state of iron pnictides are indeed from anisotropic spin excitations \cite{RMFernandes2014}, one would expect the resistivity and spin excitation anisotropy for BaFe$_2$As$_2$ to be considerably different from those of SrFe$_2$As$_2$, given the strongly first order nature of the coupled structural and magnetic phase transitions \cite{Krellner,Jesche}.
Although previous transport measurements on detwinned SrFe$_2$As$_2$ appear to bear this out \cite{SDDas}, there are no systematic studies to compare the resistivity and spin excitation anisotropy in the nearly 100\% detwinned BaFe$_2$As$_2$ and \SFA.

\begin{figure}
\includegraphics[scale=.7]{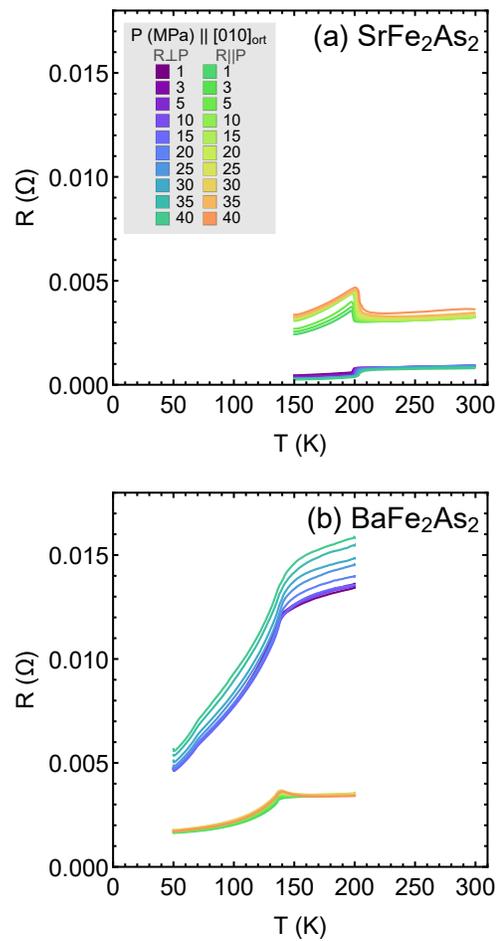}
\caption{Resistance [$R(\Omega)$] of (a) \SFA\ and (b) \BFA\ obtained using a custom built uniaxial device for all pressures and temperatures. Resistance perpendicular [$R(\Omega)\bot P$] and along [$R(\Omega)|| P$] the uniaxial pressure directions are clearly marked. 
}
\label{Resistivity-rawplot}
\end{figure}

\begin{figure*}
\includegraphics[scale=.6]{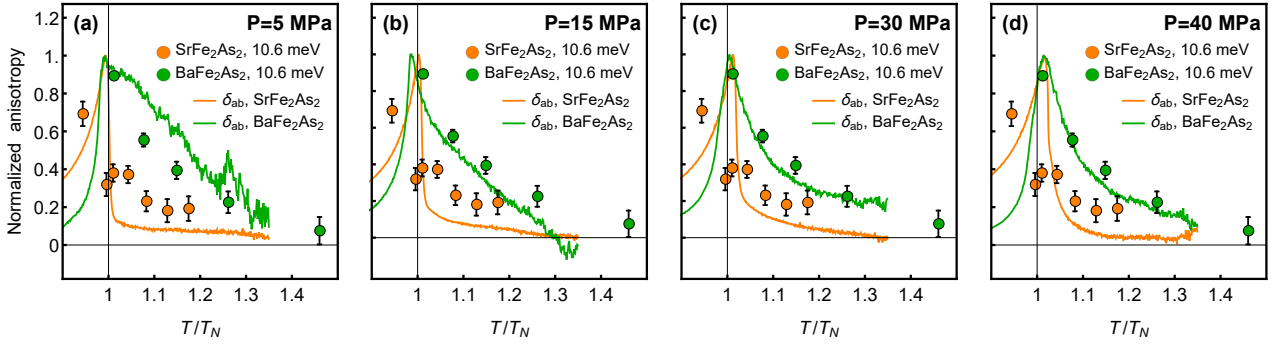}
\caption{$T/T_{\rm N}$ dependence of the resistivity anisotropy $\delta_{ab}$ at in-plane uniaxial pressures of (a) $P=5$ MPa, (b) $P=15$ MPa, (c) $P=30$ MPa, and (d) $P=40$ MPa 
for detwinned BaFe$_2$As$_2$ and SrFe$_2$As$_2$, compared to spin excitation anisotropy $(I_{10}-I_{01})/(I_{10}+I_{01})$ for detwinned BaFe$_2$As$_2$ and SrFe$_2$As$_2$, measured at $E=10.6\pm 2.8$ meV with incident neutron energy $E_i=$80 meV.
Below $P\approx$15 MPa, the finite twinning of the samples obscures the connection to spin excitation anisotropy, but for 30 and 40 MPa the connection is robust.
The data for \SFA\ show sharp changes across $T_{\rm N}$, indicative of the first order nature of the transition, and similar but broader features in \BFA.
The large differences in spin excitation anisotropy very close to $T_{\rm N}$ for these two materials may arise from their different low temperature spin anisotropy gaps \cite{Matan09,Wang13,Zhao08}. 
}
\label{fig1t}
\end{figure*}

In this paper, we report transport and inelastic neutron scattering measurements designed to study the impact of the strongly first order AF phase transition of SrFe$_2$As$_2$ on the resistivity and spin excitation anisotropy in the paramagnetic phase.
Similar to previous work \cite{xylu16,Lu2018}, we used a mechanical uniaxial pressure device to detwin multiple samples of SrFe$_2$As$_2$ for inelastic neutron scattering experiments, and compare this to resistivity measurements as a function of carefully controlled uniaxial pressure using a home-built 
instrument \cite{Tam}.
In the unstrained state, SrFe$_2$As$_2$ undergoes strongly first order coupled structural and magnetic phase transitions at $T_{\rm s}= T_{\rm N}\approx 210$ K from a paramagnetic tetragonal state to an AF orthorhombic state \cite{JZhao2008}.
Applying fixed uniaxial pressure along the orthorhombic $b$-axis, we can detwin SrFe$_2$As$_2$ below $T_{\rm s}$ and $T_{\rm N}$ for neutron scattering experiments (Fig. \ref{fig1n}).
Since we find no low-energy spin excitations at disallowed positions $\mathbf{Q}=(0, \pm1)$ [Figs. \ref{fig1n}(a) and \ref{fig1n}(c)], we conclude that the SrFe$_2$As$_2$ single crystal is nearly 100\% or completely detwinned.
Our inelastic neutron scattering experiments reveal that the spin excitation anisotropy in the paramagnetic state, defined as $(I_{10}-I_{01})/(I_{10}+I_{01})$ where $I_{10}$ and $I_{01}$ are spin excitation intensities at $\mathbf{Q}_{\rm AF} = (\pm1, 0)$ and $\mathbf{Q}=(0, \pm1)$, respectively, is dramatically different for SrFe$_2$As$_2$ and BaFe$_2$As$_2$.
In particular, the anisotropy above $T_{\rm N}$ is smaller and decays more rapidly in 
\SFA\ compared with measurements on \BFA\ under the same 
experimental conditions (Fig. \ref{Resistivity-rawplot}).
To explore the connection with the electronic nematic phase, we overlay the resistivity anisotropy in the paramagnetic state of detwinned \SFA\ and \BFA\ at several uniaxial pressures in Fig. \ref{fig1t}.
We find that the nematic phase, as revealed by pressure-induced resistivity anisotropy, also persists to a higher $T/T_{\rm N}$ in \BFA\ compared to \SFA, independent of how the data are analyzed.
Since our BaFe$_2$As$_2$ and SrFe$_2$As$_2$ single crystals are prepared the same way \cite{YCChen}, any impurity scattering in these two materials should be similar.
Since the uniaxial pressure-induced lattice distortions are similar in both materials seen in previous neutron Larmor diffraction experiments \cite{xylu16}, we conclude that the differences in the resistivity anisotropy must be the intrinsic properties of these materials.  Therefore, the resistivity anisotropy and nematic phase in the paramagnetic phase of iron pnictides are intimately associated with the nature of the magnetic phase transition and anisotropic spin excitations, consistent with expectations that the nematic phase is spin-driven \cite{RMFernandes2014}. 

\section{Experimental Results}

\subsection{Transport Measurements}
We first describe our transport measurements on detwinned BaFe$_2$As$_2$ and SrFe$_2$As$_2$ using a custom-built uniaxial detwinning instrument in a Quantum Design Dynacool physical property measurement system \cite{Tam}.
Single crystals of BaFe$_2$As$_2$ and SrFe$_2$As$_2$ were grown using the self-flux method \cite{YCChen}, aligned and cut into square shapes along the orthorhombic axes, with pressure applied along an edge (the orthorhombic $b$-axis).
The pressure is directly measured throughout the experiment using a load cell which is fed back to the controller in order to maintain constant force \cite{Tam}.
Figures  \ref{Resistivity-rawplot}(a) and \ref{Resistivity-rawplot}(b) 
show temperature dependence of 
the resistivity of SrFe$_2$As$_2$ and BaFe$_2$As$_2$, respectively. 
The resistance along the $a$- and $b$-axis directions for different values of uniaxial pressure is shown as a function of $T/T_{\rm N}$, with Ch. 1 data collected with current perpendicular to the pressure direction (Figure \ref{Resistivity-rawplot}).
The data shows somewhat different values for the four sets of measurements (two samples, two directions) due to the small size differences between the samples.
The largest source of error in the applied pressure is the estimate of the cross-sectional size of the samples, which is approximately $5.5\times 0.4 = 2.0$ mm$^2$ for \SFA\  and $4.3 \times 0.7 =  3.0$ mm$^2$ for \BFA, with approximately 10-20\% error.
In each measurement the pressure is applied at high temperature before cooling across the phase transition, and data is collected on warming at a fixed rate.
The samples were held in the uniaxial instrument between aluminum plates coated with a thin layer of Loctite E-30UT epoxy to serve as a buffer layer for even distribution of force over the sample edges.
Wires were attached near the corners of the square face to measure resistivity anisotropy by the Montgomery method \cite{HRMAN15} and the direction of current/voltage was alternated between the $a$ and $b$ axes during the course of each temperature sweep.
In this geometry, the uniaxial instrument can apply pressures between near-zero and about 150 MPa for samples of these size, enough to cover the range of pressures necessary to fully detwin the crystals ($\sim$10 MPa) and well above the pressure that causes them to break.

\begin{figure}
\includegraphics[scale=.43]{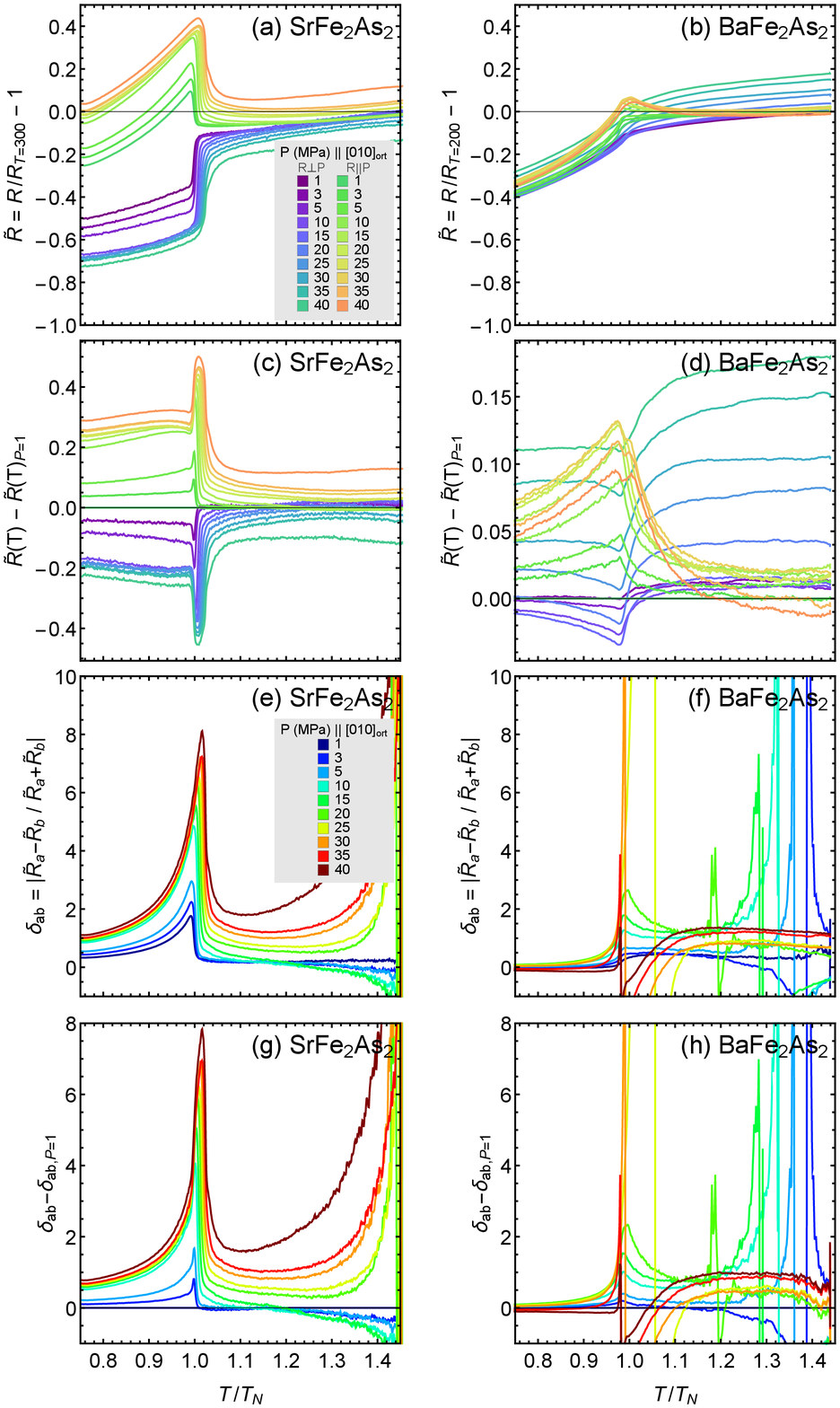}
\caption{Resistance and resistance anisotropy of \SFA\ and \BFA.
Normalized resistance $\tilde{R} = (R[T,P]-R[T_\text{max},P=1])/R[T_\text{max},P=1]$, i.e. normalized to the value at $P=1$ MPa, $T=300$ K for (a) \SFA\ or 200 K for (b) \BFA.
(c,d) Uniaxial pressure effect on the normalized resistance, $\Delta_P \tilde{R} = \tilde{R}[T,P] - \tilde{R}[T,P=1]$.
(e,f) Anisotropy $\delta_{ab}[T,P] = |\Delta_P \tilde{R}_a - \Delta_P \tilde{R}_b| /  (\Delta_P \tilde{R}_a + \Delta_P \tilde{R}_b)$ of the pressure-induced resistivity changes between $a$ and $b$ orthorhombic axes.
(g,h) Uniaxial pressure effect on the anisotropy, $\delta_{ab}[T,P] - \delta_{ab}[T,P=1]$.
}
\label{Resistivity-norm-P0-Tmax}
\end{figure}

We choose three methods of normalizing the raw data to proceed with analysis: (1) where the raw resistance data is scaled to the value at P=1 MPa and the maximum temperature (300 K for \SFA\  and 200 K for \BFA, which is approximately $1.5 T_{\rm N}$ in both cases); (2) where the raw resistance data is scaled to the value at maximum temperature ($T/T_{\rm N} \approx 1.5$) for each pressure independently; and (3) where the raw resistance data is scaled at $T/T_{\rm N}=1.2$ for each pressure. In each case, it is clear that the anisotropy persists to higher relative temperature in the case of \BFA. 
Figure \ref{fig1t} summarizes normalized temperature
$T/T_{\rm N}$ dependence of the resistivity 
anisotropy $\delta_{ab}$ under different uniaxial pressure using method (2). 
For all applied uniaxial pressures, we find that the 
resistivity anisotropy extends to larger $T/T_{\rm N}$ for \BFA\ than that of \SFA.

In Figure \ref{Resistivity-norm-P0-Tmax}, we use method (1) for normalization, which is to the value measured at the smallest pressure and the highest temperature (approximately 1.5 $T/T_{\rm N}$ in both cases).
In Figures \ref{Resistivity-norm-P0-Tmax}(a) and \ref{Resistivity-norm-P0-Tmax}(b), the first set of temperature sweeps (colors between purple and blue) show normalized resistance perpendicular to the pressure direction (the $a$-axis direction $R_a$), and the second data (green to orange) parallel to pressure (the $b$-axis direction $R_b$).
The Figures \ref{Resistivity-norm-P0-Tmax}(c) and \ref{Resistivity-norm-P0-Tmax}(d) show the same data with the lowest pressure data subtracted, yielding the intrinsic pressure effect, which for pressures less than $P=15$ MPa is a combination of detwinning and perturbative effects on the electronic structure.
Above $P=15$ MPa, we find no major qualitative changes for pressure above $\sim$15 MPa in both compounds, corresponding to complete detwinning under this pressure.
The remainder of the changes are associated with the uniaxial distortion in a single domain in the case of BaFe$_2$As$_2$ \cite{Tam} and we expect the same for SrFe$_2$As$_2$.
For example, $T_{\rm N}$ gradually shifts upward with increasing pressure, as in the case of \BFA\ \cite{Tam}, and broadens somewhat for each sample, consistent with a small distribution of uniaxial pressure over the entire sample volume.
(We note that in our data, the resistance of \BFA, Ch. 1, seems to be accumulating an offset with increasing pressure.
Since its value is constant with respect to temperature, we believe this offset is extrinsic to the sample, and in methods (2) and (3) it is automatically eliminated by the normalization.
Nevertheless, we proceed with method (1) under the assumption that it is intrinsic, for the sake of argument.)
Figures \ref{Resistivity-norm-P0-Tmax}(e) and \ref{Resistivity-norm-P0-Tmax}(f) show the absolute value of anisotropy between $R_a$ and $R_b$, $\delta_{ab}[T,P] = |\Delta_P \tilde{R}_a - \Delta_P \tilde{R}_b| /  (\Delta_P \tilde{R}_a + \Delta_P \tilde{R}_b)$.
To remove any ambiguity arising from the intrinsic temperature dependence of the resistivity, we also show in 
Figures \ref{Resistivity-norm-P0-Tmax}(g) and \ref{Resistivity-norm-P0-Tmax}(h)
the anisotropy after subtracting the anisotropy measured at the lowest pressure, $P=1$ MPa.
In principle, this extra step is not necessary since for a fully twinned crystal the anisotropy should be indistinguishable between $R_a$ and $R_b$.
The fact that the anisotropy is nonzero below $T_{\rm N}$ at only 1 MPa in both crystals may reflect the fact that the pressure is applied at high temperature before cooling across $T_s$, so even a small symmetry-breaking force can have relatively large changes on the volume fraction of different twin domains.
Since there are multiple crossing points in the pressure-subtracted data [Figures \ref{Resistivity-norm-P0-Tmax}(c) and \ref{Resistivity-norm-P0-Tmax}(d)], the anisotropy [\ref{Resistivity-norm-P0-Tmax}(e) and \ref{Resistivity-norm-P0-Tmax}(f)] contains divergence-like features in \BFA\ for most pressures where the values of $R_a$ and $R_b$ accidentally cross.
However, by comparing the relatively low-pressure data such as at 10 and 15 MPa, we can clearly see that the pressure-induced resistivity anisotropy extends to a much larger $T/T_{\rm N}$ compared with that of SrFe$_2$As$_2$, indicating that temperature regime of the nematic phase is sensitive to the first order nature of the AF phase transition in SrFe$_2$As$_2$.

\begin{figure}
\includegraphics[scale=.43]{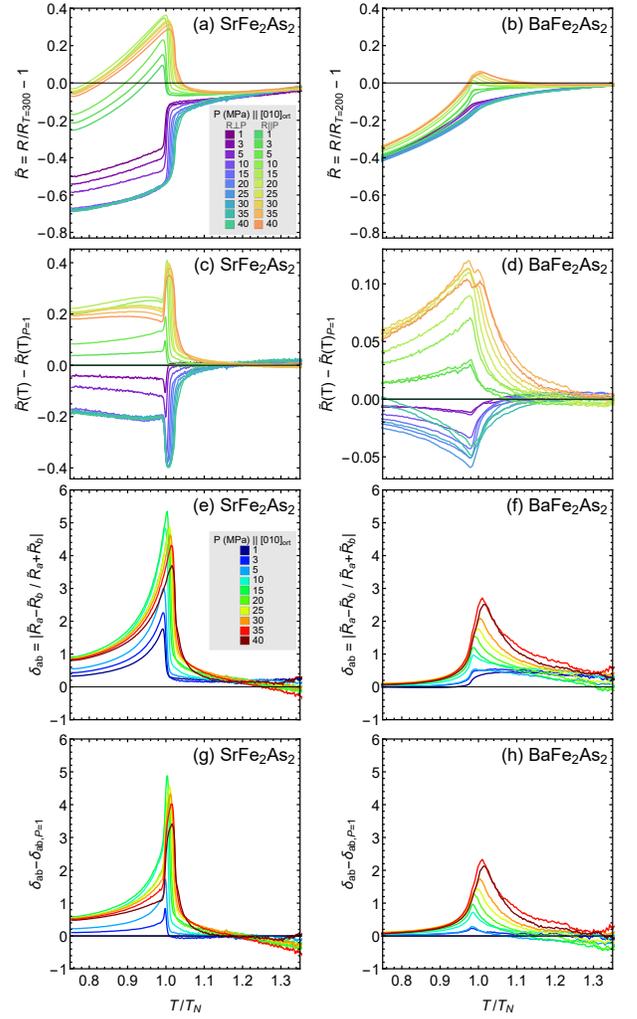}
\caption{Resistance and resistance anisotropy of \SFA\ and \BFA, with the same panels as Figure \ref{Resistivity-norm-P0-Tmax}. Here, in (a,b) we use the normalized resistance $\tilde{R} = (R[T,P]-R[T_\text{max},P])/R[T_\text{max},P]$, i.e. normalized to the value at $T=300$ K (\SFA) or 200 K (\BFA) measured at each pressure.
}
\label{Resistivity-norm-indiv-Tmax}
\end{figure}

In Figure \ref{Resistivity-norm-indiv-Tmax}, we now use method (2), normalizing each temperature sweep to its highest value (approximately 1.5 $T/T_{\rm N}$ in both cases).
This method accounts for overall changes in resistivity with increasing pressure and therefore nullifies any instrumental effects or changes caused by, for example, a small flake breaking off near one of the electrical leads.
This method most clearly shows the pressure-induced anisotropy [Figs. \ref{Resistivity-norm-indiv-Tmax}(e)-(h)] and convincingly demonstrates that the pressure effect persists to a higher $T/T_{\rm N}$ in BaFe$_2$As$_2$.

\begin{figure}
\includegraphics[scale=.43]{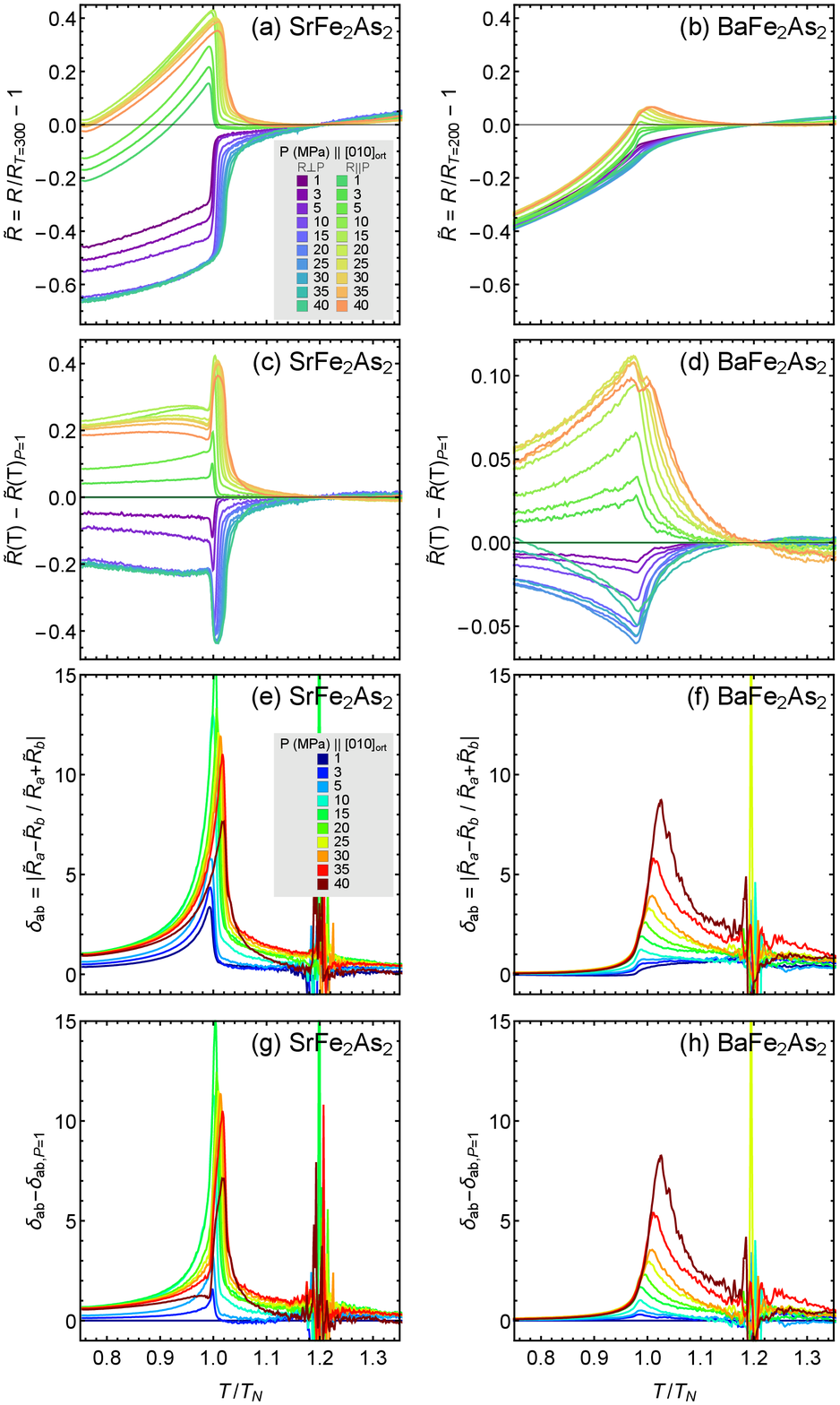}
\caption{Resistance and resistance anisotropy of \SFA\ and \BFA, with the same panels as Figure \ref{Resistivity-norm-P0-Tmax}. Here, in (a,b) we use the normalized resistance $\tilde{R} = (R[T,P]-R[T=1.2T_N,P])/R[T=1.2T_N,P]$, i.e. normalized to the value at $T=238$ K (\SFA) or 166 K (\BFA) measured at each pressure.
}
\label{Resistivity-norm-indiv-T1p2}
\end{figure}

Finally, in Figure \ref{Resistivity-norm-indiv-T1p2}, we use method (3), normalizing to the value at $T=1.2 T_{\rm N}$ in each temperature sweep. This accounts for any possible differences in anisotropy between the compounds that may be related to temperature-induced disorder effects. Nevertheless, we recover the same conclusion that the anisotropy decays more rapidly in \SFA\ compared to \BFA.

We make particular note here about the values chosen for $T_N$, since it has a measurable impact on the data in this case. In particular, we find a lower $T_N \sim 199$ K for \SFA\ compared with the values shown in the main text from neutron scattering $T_N \sim 210$ K. We believe the values chosen are correct in both cases, and that the uniaxial pressure is actually much higher in the samples used for neutron scattering such that the ordering temperature is increased by about 10 K. The increase in ordering temperature under pressure is a well-known effect, and is seen clearly in the present data. 

Finally, we point out we have not normalized the anisotropy in these figures, except in Figure \ref{fig1t}.
We believe the non-normalized anisotropy is a good measure of the intrinsic resistivity anisotropy under constant strain, since the lattice anisotropy under 30 MPa is known to be similar between \SFA\ and \BFA\ \cite{xylu16}.

 
%
\begin{figure}
\includegraphics[scale=.4]{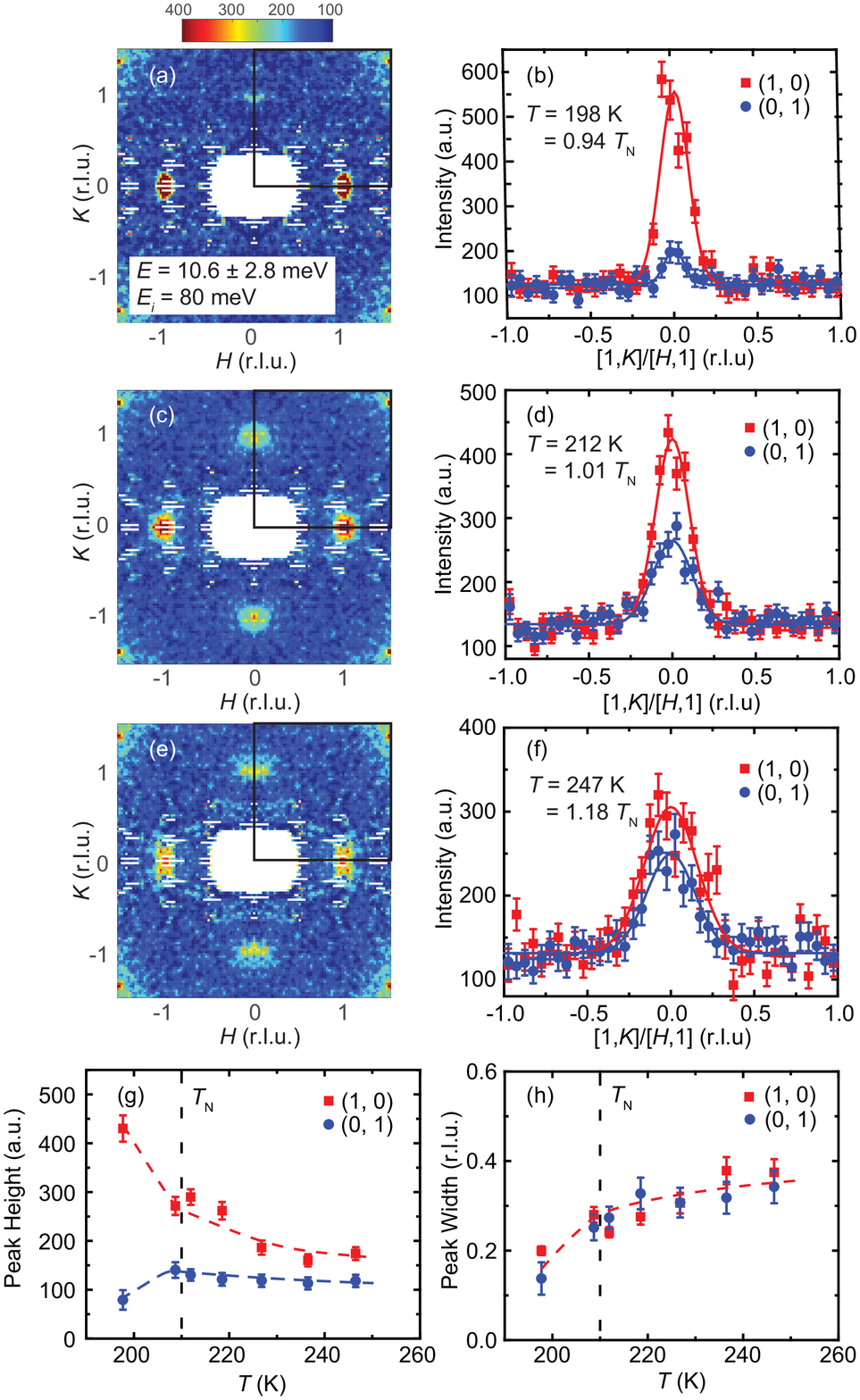}
\caption{Temperature dependence of the spin excitation anisotropy as a function
of increasing temperature across $T_{\rm N}$ for SrFe$_2$As$_2$. (a,b) Spin waves
of an energy transfer $E=10.6\pm 2.8$ meV at $T=0.94T_{\rm N}$ in the $[H,K]$ plane, and cuts along
the $[H,1]$ and $[1,K]$ directions. Identical scans at (c,d) $T=1.01T_{\rm N}$,
and (e,f) $T=1.18T_{\rm N}$. (g) Temperature dependence of the peak intensity
at $I_{10}$ and $I_{01}$ across $T_{\rm N}$. (h) Temperature dependent width of 
spin excitations across $T_{\rm N}$ for SrFe$_2$As$_2$. }
\label{fig3}
\end{figure}
%

\subsection{Inelastic Neutron Scattering Measurements}

To see if the temperature dependence of the spin excitation anisotropy in SrFe$_2$As$_2$ 
follows that of the resistivity anisotropy, we measured low-energy spin excitations
across $T_{\rm N}$ with inelastic neutron scattering experiments performed at the MERLIN time-of-flight neutron-scattering spectrometer at ISIS, Rutherford Appleton Laboratory \cite{MERLIN}. The single crystals were detwinned under uniaxial pressures of at least $30$ MPa. We define the wave vector \textbf{Q} in three-dimensional reciprocal space in \AA$^{-1}$ as ${\bf Q}=H {\bf a^\ast}+K{\bf b^\ast}+L{\bf c^\ast}$, where $H$, $K$, and $L$ are Miller indices and ${\bf a^\ast}=\hat{{\bf a}}2\pi/a, {\bf b^\ast}=\hat{{\bf b}} 2\pi/b, {\bf c^\ast}=\hat{{\bf c}}2\pi/c$ are reciprocal lattice units (r.l.u.) [Fig. 1(b)]. In the low-temperature AF orthorhombic phase of SrFe$_2$As$_2$, $a\approx 5.57$ \AA, $b\approx 5.51$ \AA, and $c\approx 12.29$ {\AA} \cite{JZhao2008}. The sample array was aligned with the $c$-axis along the incident beam direction (${\bf k_i}\parallel c$) with neutron energy of $E_i=80$ meV. We carried out measurements
at many temperatures above and below $T_{\rm N}$ to obtain temperature dependence of $I_{10}$ and $I_{01}$ for comparison with BaFe$_2$As$_2$ \cite{Lu2018}.

In the AF ordered state,  
spin waves from the collinear AF order should stem from $\mathbf{Q}_{\rm AF} = (\pm1, 0)$ with $L=\pm1, \pm 3$ in reciprocal space [Fig. 1(b)] \cite{Lu14}. On warming to the paramagnetic phase, 
the scattering should have very weak $L$-dependence \cite{Lu2018}.
Figures \ref{fig3}(a) and \ref{fig3}(b) show spin waves of energy transfer $E=10.6\pm 2.8$ meV
and the corresponding cuts along the $[H,1]/[1,K]$ directions  
at $T=0.94 T_{\rm N}$. As expected, we find spin waves at $\mathbf{Q}_{\rm AF} = (\pm1, 0)$ dominating the
scattering and very weak magnetic scattering at $(0,\pm 1)$, 
 consistent with the nearly 100\% detwinning ratio shown in Fig. 1(c).  
On warming to $T=1.01 T_{\rm N}$, we see a significant reduction in the spin excitation
anisotropy at  $\mathbf{Q}_{\rm AF} = (\pm1, 0)$ ($I_{10}$) 
and $(0,\pm 1)$ ($I_{01}$) [Figs. \ref{fig3}(c) and \ref{fig3}(d)].  On further warming to $T=1.18 T_{\rm N}$,
we find that spin excitations at  $\mathbf{Q}_{\rm AF} = (\pm1, 0)$ 
and $(0,\pm 1)$ almost become equal in intensity, but have weak temperature dependence. 
These results thus suggest that the remaining spin excitation
anisotropy is due to the presence of uniaxial pressure [Figs. \ref{fig3}(e) and \ref{fig3}(f)].  Figure \ref{fig3}(g) shows  temperature dependence of the spin excitation 
intensity at $\mathbf{Q}_{\rm AF} = (\pm1, 0)$ ($I_{10}$) and $(0,\pm 1)$ ($I_{01}$).  
Since the widths of the spin excitations change smoothly across $T_{\rm N}$ as shown in Fig. \ref{fig3}(h),
we conclude that the spin excitation anisotropy at $E=10.6\pm 2.8$ meV 
reduces dramatically across $T_{\rm N}$.

%
\begin{figure}
\includegraphics[scale=.4]{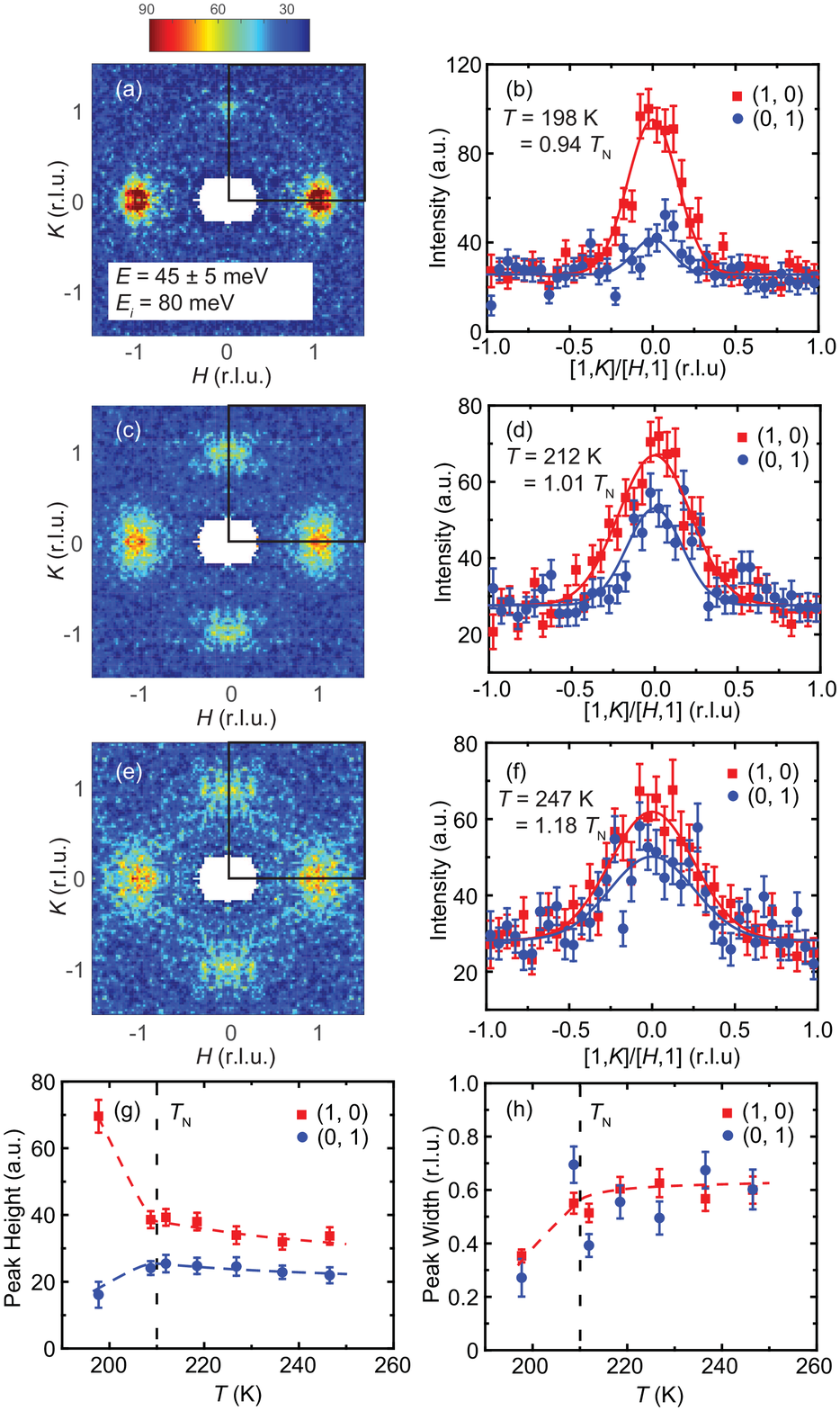}
\caption{Temperature dependence of the spin excitation anisotropy at an energy 
transfer of $E=45\pm 5$ meV as a function
of increasing temperature across $T_{\rm N}$ for SrFe$_2$As$_2$. (a,b) Spin waves
of  $E=45\pm 5$ meV at $T=0.94T_{\rm N}$ in the $[H,K]$ plane, and cuts along
the $[H,1]$ and $[1,K]$ directions. The excitation anisotropy is somewhat smaller than that
at $E=10.6\pm 2.8$ meV. Identical scans at (c,d) $T=1.01T_{\rm N}$,
and (e,f) $T=1.18T_{\rm N}$. (g) Temperature dependence of the peak intensity
at $I_{10}$ and $I_{01}$ across $T_{\rm N}$. The persistent spin excitation anisotropy above
$T_{\rm N}$ is due to the presence of uniaxial pressure, similar features 
are also seen in BaFe$_2$As$_2$. 
(h) Temperature dependent width of 
spin excitations across $T_{\rm N}$ for SrFe$_2$As$_2$.}
\label{fig4}
\end{figure}
%

Having examined the spin excitation anisotropy above $T_{\rm N}$ at low energies just above the spin wave gap, we turn to the energy dependence at a higher energy transfer $E=45\pm 5$ meV, which is about 25\% of the total magnetic bandwidth.  At  $T=0.94 T_{\rm N}$, we also see clear spin wave anisotropy
with most of the spectral weight at $\mathbf{Q}_{\rm AF} = (\pm1, 0)$, quite similar to spin waves 
at $E=10.6\pm 2.8$ meV [Figs. \ref{fig4}(a) and \ref{fig4}(b)]. On warming to $T=1.01 T_{\rm N}$, spin excitations
at $\mathbf{Q}_{\rm AF} = (\pm1, 0)$ and $(0,\pm 1)$ are still anisotropic [Figs. \ref{fig4}(c) and \ref{fig4}(d)], but much less so compared with data at  $E=10.6\pm 2.8$ meV [Figs. \ref{fig3}(c) and \ref{fig3}(d)].  Finally, we see very little spin
excitation anisotropy at $T=1.18 T_{\rm N}$ [Figs. \ref{fig4}(e) and \ref{fig4}(f)], very similar to the data
at  $E=10.6\pm 2.8$ meV [Figs. \ref{fig3}(e) and \ref{fig3}(f)].  Figures \ref{fig4}(g) and \ref{fig4}(h) show the temperature
dependence of the magnetic scattering intensity and width of the spin excitations, respectively,  
at $\mathbf{Q}_{\rm AF} = (\pm1, 0)$ and $(0,\pm 1)$.

To quantitatively summarize the spin excitation 
anisotropy in the paramagnetic state of 
SrFe$_2$As$_2$ and BaFe$_2$As$_2$, we plot in Figure \ref{fig1t} the 
relative temperature ($T/T_{\rm N}$) dependence of the spin excitation anisotropy at low energy for these two materials under nearly 100\% detwinning,
next to the resistivity anisotropy measured on our uniaxial instrument.
In the paramagnetic state, we see a clear difference in the temperature dependence of the spin excitation
anisotropy at an energy transfer $E=10.6\pm 2.8$ meV, where the spin excitation anisotropy for BaFe$_2$As$_2$ extends to much
higher $T/T_{\rm N}$ than that of SrFe$_2$As$_2$. These results are qualitatively consistent with 
transport measurements of the resistivity anisotropy for SrFe$_2$As$_2$ and BaFe$_2$As$_2$.

\section{Discussion and Conclusions}

Theoretically, the electronic nematic phase and associated resistivity anisotropy is expected to only occur below
the tetragonal-to-orthorhombic phase transition temperature $T_{\rm s}$ \cite{RMFernandes2014}.  Although recent neutron pair distribution function and Lamor diffraction experiments
on different classes of iron pnictides including Sr$_{1-x}$Na$_x$Fe$_2$As$_2$ 
\cite{Frandsen18} and NaFe$_{1-x}$Ni$_x$As \cite{WYWang18} reveal clear evidence for local
orthorhombic lattice distortions in temperatures well above $T_s$, these local lattice distortions are evenly distributed along the two orthorhombic lattice directions and therefore not expected to induce resistivity anisotropy. 
The clear presence of resistivity \cite{JHChu2010,JHChu2012}, spin excitation 
\cite{Lu14,HRMAN15,Wenliang,HRMan18}, 
and orbital population anisotropy \cite{MingYi17} in the paramagnetic phase of iron pnictides can arise from 
the interaction of applied uniaxial pressure with nematic susceptibility and associated spin excitations  
through magnetoelastic coupling \cite{Yli18}. The applied uniaxial pressure 
should be mostly sensitive to low energy spin excitations and acoustic phonons, and have little
impact to high energy spin excitations.
Since uniaxial pressure applied on the system has already broken 
the tetragonal symmetry of the paramagnetic phase, the system can only exhibit a paramagnetic to AF phase transition below $T_{\rm N}$ \cite{xylu16}.
 If nematic order in the paramgnetic phase of iron pnictides is driven by spin fluctuations associated with the static AF order \cite{Karahasanovic,Christensen2016}, one would expect that the nature 
of the magnetic phase transition will affect critical spin fluctuations 
in the paramagnetic state near $T_{\rm N}$. 
For systems with a strongly first order AF phase transition, 
for instance SrFe$_2$As$_2$ \cite{Krellner,Jesche}, one would 
expect weak or no critical spin fluctuations associated with the magnetic order in the paramagnetic state. 
On the other hand, the AF phase transition in BaFe$_2$As$_2$ is a weakly first order transition, and
doping Co and Ni as well as uniaxial pressure drive the system into a second order AF phase transition \cite{Lester09,Nandi,XYLu13}. 
Therefore, one would expect to find an extended critical regime and considerable critical magnetic scattering in the paramagnetic state.

By comparing the temperature dependence of the low energy spin excitation anisotropy of BaFe$_2$As$_2$ and
SrFe$_2$As$_2$, we see a much faster reduction in spin excitation anisotropy in SrFe$_2$As$_2$, which is consistent with our transport measurements.  Our experiments therefore establish a direct correlation between critical spin excitations in the paramagnetic state and resistivity anisotropy. Using the same reasoning, we would expect weak resistivity anisotropy and 
electronic nematic phase
in the paramagnetic tetragonal phase of 
hole-doped iron pnictides Ba$_{1-x}$K$_x$Fe$_2$As$_2$ \cite{Avci12} and isoelectronic doped
BaFe$_{2}$(As$_{1-x}$P$_x$)$_2$ \cite{Allred,Hu2015}, since both materials have coupled first order
structural and magnetic phase transitions. Indeed, transport measurements  
on uniaxial pressure detwinned Ba$_{1-x}$K$_x$Fe$_2$As$_2$ reveal a much smaller
region of resistivity anisotropy above $T_{\rm N}$ compared with similarly prepared electron-doped
 BaFe$_{2-x}$Co$_x$As$_{2}$ \cite{JJYing}. On the other hand, since 
 annealing as-grown single crystals of 
 BaFe$_2$As$_2$ improves the first order nature of the magnetic phase transition \cite{Rotundu10}, reduces the disorder, resistivity anisotropy, and magnitude 
of residual resistivity \cite{Ishida11,Ishida13}, one would expect reduced spin excitation anisotropy in the paramagnetic phase of annealed BaFe$_2$As$_2$. It would therefore interesting to carry out
studies of the annealing effect on the spin excitation anisotropy of BaFe$_2$As$_2$.

In summary, we have used transport and inelastic neutron scattering 
to study the effect of a strongly first order magnetic phase transition on the 
magnitude and temperature dependence of resistivity and spin excitation anisotropy 
in the paramagnetic phase of SrFe$_2$As$_2$ and BaFe$_2$As$_2$. We find that the resistivity
and spin excitation anisotropy in the paramagnetic state of iron pnictides are highly dependent on the nature of the magnetic phase transition.  For SrFe$_2$As$_2$, a system with a first order
magnetic phase transition, both the resistivity and spin excitation anisotropy disappear rapidly
in the paramagnetic phase close to $T/T_{\rm N}$ due to a lack of critical spin fluctuations. For BaFe$_2$As$_2$, a system with a weakly first order or second order phase transition, the resistivity and spin excitation anisotropy 
extend to much higher $T/T_{\rm N}$.  These results are consistent with expectations of a  
spin excitation driven electronic nematic phase in the paramagnetic phase of iron pnictides, providing further evidence for the importance of magnetism to the electronic properties and superconductivity of iron based superconductors.

\section{Acknowledgments}

The neutron-scattering
work at Rice University was supported by the US NSF Grant
No. DMR-1700081 (P.D.). The SrFe$_2$As$_2$ single-crystal
synthesis work at Rice University was supported by the Robert
A. Welch Foundation Grant No. C-1839 (P.D.). Li Zhang was supported by the 
Natural Science Foundation of China (No. 61376094) and China Scholarship 
Council (No.201408330028).



\end{document}